\documentclass[prl,aps,reprint,superscriptaddress]{revtex4-1}
\usepackage[latin1]{inputenc} % unicode lib and save format UTF-8 for text files +  umlaut
\usepackage[T1]{fontenc} % text encoding
\usepackage[english]{babel} % correct language settings
\usepackage{amsmath,amssymb} % math packages
\usepackage{graphicx,color} % graphic and color

\usepackage[breaklinks,colorlinks,linkcolor=blue,citecolor=blue,urlcolor=blue]{hyperref}
\newcommand{\er}[1]{Eq.~\eqref{#1}}
\newcommand{\ers}[2]{Eqs.~(\ref{#1},\ref{#2})}
\newcommand{\pd}[1]{\partial_{#1}}

\newcommand{\expec}[1]{\langle\!\langle #1 \rangle\!\rangle}

\newcommand{\im}[0]{\mathrm{i}}
\newcommand{\Or}[1]{\mathcal{O}(#1)}

\newcommand{\Ln}[0]{\mathcal{L}_\mathrm{n}}
\newcommand{\Lphi}[0]{\mathcal{L}_{\phi}}
\newcommand{\Ls}[0]{\mathcal{L}_\mathrm{s}}
\newcommand{\twist}[0]{\vec{A}_\theta}

\begin{document}

\title{Superfluid Stiffness of a Driven Dissipative Condensate with Disorder}

\author{Alexander Janot}
\affiliation{Institut f\"ur Theoretische Physik, Universit\"at Leipzig, 04009 Leipzig, Germany}
\author{Timo Hyart}
\affiliation{Instituut-Lorentz, Universiteit Leiden, P.O. Box 9506, 2300 RA Leiden, The Netherlands}
\author{Paul R. Eastham}
\email[Corresponding author: ]{easthamp@tcd.ie}
\affiliation{School of Physics and CRANN, Trinity College, Dublin 2, Ireland}
\author{Bernd Rosenow}
\affiliation{Institut f\"ur Theoretische Physik, Universit\"at Leipzig, 04009 Leipzig, Germany}

\begin{abstract}
Observations of macroscopic quantum coherence in driven systems, e.g. polariton condensates, have strongly stimulated experimental as well as theoretical efforts during the last decade. We address the question of whether a driven quantum condensate is a superfluid, allowing for the effects of disorder and its non-equilibrium nature. We predict that for spatial dimensions $d < 4$ the superfluid stiffness vanishes once the condensate exceeds a critical size, and treat in detail the case $d=2$. Thus a non-equilibrium condensate is not a superfluid in the thermodynamic limit, even for weak disorder, although superfluid behavior would persist in small systems. 
\end{abstract}
\maketitle
%

%%% Introduction %%%
Perhaps the most spectacular manifestation of Bose-Einstein condensation, and its associated macroscopic quantum coherence, is superfluidity. Recent experiments~\cite{Kasprzak:2006} have shown macroscopic quantum coherence in a population of mixed matter-light excitations, so-called polaritons (see~\cite{Deng:2010} for a review). Aspects of superfluid behavior, including quantized vortices~\cite{Lagoudakis:2008,Sanvitto:2010} and suppression of scattering from defects~\cite{Amo:2009}, have also been observed. However, unlike the constituents of conventional condensates, such as cold atoms, polaritons have a finite lifetime. Thus, the polariton condensate is a non-equilibrium steady-state, in which the losses are compensated by particles flowing in from an external source. This leads to the interesting possibility of new universal behavior, different from that found in equilibrium~\cite{Sieberer:2013}.  Many similarities, nonetheless, appear to remain, at least in the absence of disorder: perturbatively, the forms of the correlation functions are the same as in equilibrium~\cite{Chiocchetta:2013,Roumpos:2012} (long-range order in three dimensions, and quasi-long-range order in two); superfluidity is predicted to survive~\cite{Keeling:2011} ($d\ge 2$); and the static behavior, in three dimensions, involves the standard O(2) critical exponents~\cite{Sieberer:2013}.  A new dynamical critical exponent has, however, recently been discovered~\cite{Sieberer:2013}.

In equilibrium, as predicted in a seminal work~\cite{Fisher:1989}, the presence of sufficiently strong disorder may suppress the superfluid state, and cause a transition to the Bose glass. Neglecting the gain and loss processes, a similar transition to a glass-like state was predicted~ \cite{Malpuech:2007} for polaritons. Here we show, however, that these non-equilibrium processes play a fundamental role. We consider the experimentally relevant case of two dimensions, and show that, for a driven open condensate, static disorder destroys long-range order. Furthermore, the superfluid stiffness, as probed by the energy shift induced by twisted boundary conditions~\cite {Fisher:1973}, vanishes in the thermodynamic limit. Thus a two-dimensional driven condensate is not formally a superfluid, except for zero disorder, although superfluid behavior would persist below a critical length scale. We identify this length scale, and the mechanism responsible for the destruction of superfluidity, below. Our results have implications both in the search for superfluidity in polariton condensates~\cite{Lagoudakis:2008,Sanvitto:2010,Amo:2009,Keeling:2011}, and in the emerging study of non-equilibrium phase transitions in quantum many-body systems~\cite{Sieberer:2013,Chiocchetta:2013,Roumpos:2012}. Experiments on polariton condensates may involve a significant level of static disorder~ \cite{Helena:2012}, and it is therefore important to establish how disorder affects a driven condensate. 

%%% Model %%%
A phenomenological description of the macroscopic wave-function $\Psi(\vec{x},t)$ of a weakly interacting Bose condensate with gain and loss is the extended Gross-Pitaevskii Equation (eGPE)~\cite{Wouters:2007,Keeling:2008},
\begin{align}
	 \label{eq:eGPE}
	\im \hbar \pd{t} \Psi =  \left( -J \nabla^2 + V(\vec{x}) + U \left|\Psi\right|^2 \right)  \Psi + \im \left(\gamma -  \Gamma \left|\Psi\right|^2 \right) \Psi \ ,
\end{align}
where $J = \hbar^2/2m$, $V$ is a random  potential, and $U>0$ the interaction strength. The second term on the right introduces driving and losses, with $\gamma/\hbar$ the net linear gain, i.e. the stimulated in-scattering rate minus the loss rate, and a nonlinearity with gain depletion parameter $\Gamma$ (see~\cite{Keeling:2008}). These terms balance for a condensate density $n_0\equiv\gamma/\Gamma$. For $V$ we choose $\delta$-correlated Gaussian disorder
\begin{displaymath}
	\expec{V(\vec{x})} = 0 \ , \quad \expec{V(\vec{x})V(\vec{y})} = V_0^2 \delta^{(d)}(\vec{x}-\vec{y}) \ ,
\end{displaymath}
with strength $V_0$; $\expec{\ldots}$ denotes the disorder average. It is convenient to introduce units of length, time, and energy, namely, the healing length $\xi \equiv \sqrt{J/n_0U}$, $\hbar/n_0U$, and the blue shift $n_0U$, respectively. We define a dimensionless wave-function $\psi \equiv \Psi/\sqrt{n_0}$, a disorder potential $\vartheta(\vec{x}) \equiv V(\vec{x})/n_0U$ with strength $\kappa $, and a non-equilibrium control parameter $\alpha$, such that $\alpha = 0$ in equilibrium. These parameters are
\begin{align}
	\kappa \equiv \frac{V_0}{\xi^{d/2}\:n_0U} \ , \qquad \alpha \equiv \frac{\Gamma}{U} \ .
\end{align}
In the following we consider steady-state solutions of \er{eq:eGPE}. Then,  the polariton condensate emits coherent light of one frequency $\omega$, and has a time-independent density (in contrast to a desynchronized regime~\cite{Wouters:2008,Eastham:2008} with several frequencies).
With the ansatz
\begin{align}
	\label{eq:ssansatz}
	\psi(\vec{x},t) = \sqrt{n(\vec{x}\,)}\: e^{\im\phi(\vec{x}\,) - \im \omega t} \ ,
\end{align}
we obtain coupled differential equations for the condensate density $n$ and current $n\nabla\phi$,
\begin{align}
	\label{eq:hydro1}
	\omega &= (\nabla \phi)^2 + \frac{1}{4} \frac{(\nabla n)^2}{n^2} - \frac{1}{2}  \frac{\nabla^2 n}{n} + n + \vartheta \ , \\
	\label{eq:hydro2}
	0 &= \nabla \cdot (n\nabla \phi) + \alpha\ n(n-1) \ .
\end{align}
\er{eq:hydro1} determines the condensate emission frequency (chemical potential), $\omega$, while \er {eq:hydro2} is a non-equilibrium continuity equation, taking into account the coupling of the driving and losses to condensate currents. Thus, regions with $n(\vec{x}) < 1$ and $n(\vec{x}) > 1$ act as local sources and sinks, respectively. Since there is no net current through the boundary, the first term in \er {eq:hydro2} vanishes when integrated over space, while the second gives the constraint
\begin{align}
	\label{eq:constraint}
	\bar{n} \equiv \frac{1}{\Omega} \int_{\vec{x}} n(\vec{x})  = \frac{1}{\Omega} \int_ {\vec{x}} n(\vec{x})^2 \ ,
\end{align}
where $\Omega=L^d$ is the system volume (area).

As pointed out elsewhere~\cite{Wouters2:2010}, the application of the Landau criterion to a driven condensate gives a vanishing critical velocity. Nonetheless, for the clean system superfluidity has been shown to survive~\cite{Keeling:2011}, if it is defined by the irrotational current response at long wavelengths~\cite{Nozieres:1966}. We therefore probe superfluidity in the disordered case by 
%considering the response to an appropriate perturbation.  In particular, we 
applying a twist of the phase $\phi_\theta(\vec{x} + L\: \vec{e}_\theta) - \phi_\theta(\vec{x}) = \theta$ between two boundaries of the condensate separated by its size $L$ in the direction $\vec{e}_\theta$. This is equivalent to a local transformation $\nabla \phi_\theta = \nabla \phi + \twist$ where $\twist \equiv (\theta/L) \vec{e}_\theta$ is the twist current and $\phi(\vec{x})$ satisfies periodic boundary conditions.  The superfluid stiffness is then~\cite{Leggett:1970,Fisher:1973}
\begin{align}
	\label{eq:stiffness}
	f_s = \lim \limits_{\theta \to 0}\ \frac{L^2}{\theta^2}[\omega(\theta) - \omega(0)].
\end{align}

%%% Weak disorder limit -- analytic results %%%
In the limit of weak disorder,  we perturbatively solve  \ers{eq:hydro1}{eq:hydro2}
by expanding  the fields $n, \nabla \phi$, and the frequency $\omega$  in
powers of $\kappa$: $n = 1+\eta_{(1)}+\Or{\kappa^2}$ and  $\nabla \phi =  \nabla \phi_{(1)}+\Or{\kappa^2}$ with $\eta_{(1)},\nabla \phi_{(1)} \sim \Or{\kappa}$.  All disorder contributions for the frequency are of even order in $\kappa$.
This approach does not, in general, account for vortex formation~\cite {Keeling:2008,Wouters3:2010}.
To confirm that vortices can indeed be neglected,  we have performed direct numerical simulations of \er{eq:eGPE} starting from initial conditions both with and without vortices. We find that dynamically stable, well separated vortex-antivortex pairs do, in some parameter regimes, occur, but they always significantly increase the frequency of the condensate. We consider the low-energy sector, which will be selected by thermalization processes at low temperature, and focus on solutions without vortices where the circulation, $\oint \cdot \nabla \phi = 0$, vanishes around any closed path.
The leading order solution of \ers{eq:hydro1}{eq:hydro2} with $\twist\neq0$ in momentum space is
\begin{align}
	\label{eq:eta1}
	\eta_{(1)}(k) &= G_\eta(k,\twist) \vartheta_k \ ,\\
	\label{eq:phi1}
	\phi_{(1)}(\vec{k}) &= G_\phi(k,\twist) \vartheta_k \ ,
\end{align}
with
\begin{align}
	\label{eq:densgreens}
	G_\eta(k,\twist) =  \frac{-k^2 \chi_k}{k^2+2\: \im\vec{k}\cdot\twist(\im\vec{k}\cdot\twist+\alpha) \chi_k} \ ,\\
	\label{eq:phasegreens}
	G_\phi(k,\twist) =\frac{-(\im\vec{k}\cdot\twist+\alpha) \chi_k}{k^2+2\: \im\vec{k} \cdot\twist(\im\vec{k}\cdot\twist+ \alpha) \chi_k} \ ,
\end{align}
and response function $\chi_k \equiv (k^2/2+1)^{-1}$. We point out that this steady-state is a stable fixed point of the dynamical system, since the excitation spectrum of a driven condensate is diffusive~\cite{Wouters:2006,Szymanska:2006,Wouters:2009}, i.e., has both real and imaginary parts. The latter leads to an exponential decay in time for any excitation. The condensate frequency, up to quadratic order in $\kappa$, is
\begin{align}
	\label{eq:omega2}
	\expec{\omega} \approx 1+\twist^{\:2}+\int_{\vec{k}} \left\{ k^2 \left(|G_\phi|^2  -  \frac{1}{4}|G_\eta|^2 \right) - | G_\eta|^2 \right\}\kappa^2 \ .
\end{align}
Here the second order density fluctuations $\expec{\eta_{(2)}}$ were calculated using \er{eq:constraint}. Since the condensate phase is a massless mode for $\twist = 0$, its propagator behaves like $G_\phi \sim k^{-2}$ at long wavelengths, 
% In perturbation theory this 
leading to infrared divergences of the momentum integrals which we regularize by a finite-size cut-off at the wavevector $2\pi/L$. Note that any non-zero average of the disorder potential, $\bar{\vartheta} \neq 0$, can be compensated by a shift of the frequency $\omega$, see \er{eq:hydro1}. Thus we may take $\vartheta_k |_{k=0}=0$, implying $\eta_{(1)}(k),\nabla\phi_{(1)}(k)|_{k=0}=0$.

In the following we consider $d=2$ dimensions and, first, discuss the ground state properties, $\twist=0$. Using  Eqs. (\ref{eq:eta1}) and (\ref{eq:densgreens}), one finds  that the correlation function for density fluctuations decays exponentially, with the healing length $\xi$ as the decay length. Thus, density fluctuations tend to screen the disorder potential, largely uninfluenced by the driving mechanism. As discussed in Ref.~\cite{Nattermann:2008}, also significant is the density Larkin length $\Ln\sim 1/\kappa$,   at which the energy cost of density fluctuations balances the energy gained from collective pinning in the random potential; in equilibrium, superfluidity occurs for $\Ln \gtrsim 1$~\cite{Nattermann:2008}, see Eq. (\ref{eq:fs}).  A strong effect of the driving appears through the result for the phase correlation function, Eqs. (\ref{eq:phi1}) and (\ref{eq:phasegreens}). In particular, density fluctuations generate random sources and sinks, and hence random currents, causing the phase to fluctuate and  destroying long-range order in the wavefunction
\begin{displaymath}
	\expec{\psi^*(\vec{x})\psi(\vec{0})}\approx   e^{-\frac{1}{2}\expec{[\phi(\vec{x})-\phi(\vec{0})]^2}} \sim   \exp(-\vec{x}^{\:2} / \Lphi^2) \ .
\end{displaymath}
Here,   sub-leading contributions from density fluctuations and logarithmic finite-size corrections were neglected. The phase correlation length is $\Lphi\sim 1/\alpha\kappa$, defined such that the typical phase variation over this distance is of order $2\pi$. This scale can also be obtained by a generalized Imry-Ma analysis~\cite{ImryMa:1975}. We integrate Eq. (\ref{eq:hydro2}) over a region of linear size $\Lphi$: the first term becomes the current through the region's boundary, of order $\Lphi\nabla\phi\sim 1$, which accounts for the non-equilibrium current generated according to the second term, of order $\alpha\kappa\sqrt{(\Lphi/\xi)^2}$ (since $\eta_1\sim\kappa$ at scale $\xi$). As was recently also found for the driven Jaynes-Cummings-Hubbard model~\cite{Kulaitis:2013}, driving and potential disorder combine to act as phase disorder, destroying long-range order according to an Imry-Ma analysis.
In the next step we calculate the condensate stiffness using \er{eq:stiffness}, perturbatively to order $\kappa^2$,
\begin{align}
	\label{eq:fs}
	f_s \approx 1- \left\{ c_1 + g_1(L)\: \alpha^2 + \left( g_2(L) + c_2 L^2 \right)  \alpha^4 \right\} \kappa^2 \ ,
\end{align}
where we have omitted finite-size corrections vanishing for $L\to\infty$. The coefficients in this expansion are
\begin{align}
	c_1 &= \frac{1}{2\pi} \ , \qquad c_2 =  \frac{1}{(2\pi)^3} \ , \nonumber \\
	g_1(L) &= - \frac{1}{\pi} \left( \log \frac{2L^2}{(2\pi)^2} - \frac{19}{12} \right) ,  \nonumber \\
	g_2(L) &= - \frac{1}{\pi} \left( \log \frac{2L^2}{(2\pi)^2} - \frac{13}{12} \right) . \nonumber
\end{align}

In the equilibrium limit, $\alpha \to 0$, \er{eq:fs} reproduces previous findings~ \cite{Huang:1992,Meng:1994,Giorgini:1994}. As the disorder strength, $\kappa$, increases from zero the stiffness continuously reduces, before vanishing at the critical strength $\kappa=\sqrt{2\pi}$. In contrast, for a driven condensate, the perturbative result breaks down in the thermodynamic limit $L\to\infty$, for any non-zero disorder strength. We observe that the fastest divergence is controlled by the length scale $\Ls \sim 1/\alpha^2 \kappa$, and below this scale the perturbative result remains finite and physical. Thus, for systems smaller than $\Ls$ we expect superfluid behavior; however a driven disordered condensate is not a superfluid in the thermodynamic limit. 
Generalizing~\er{eq:fs} to arbitrary dimensions $d$, we find a suppression of superfluidity proportional to $L^{4-d}$ and, thus, expect that superfluidity is destroyed for  all $d <4$.

%%% numeric results %%%
%
\begin{figure}[tb]
	\centering
	\includegraphics[width=0.48\hsize]{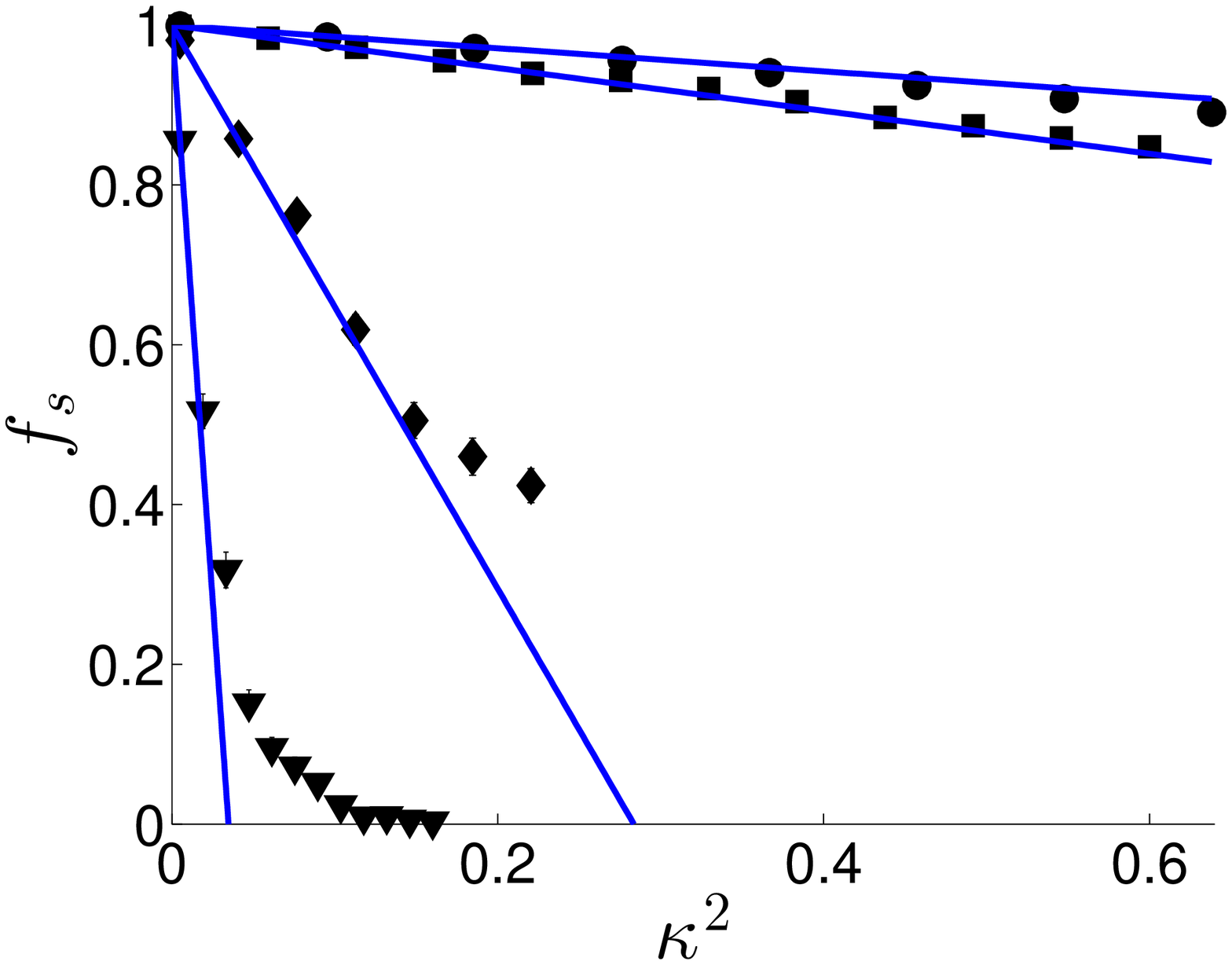} \nolinebreak \hfill
	\includegraphics[width=0.48\hsize]{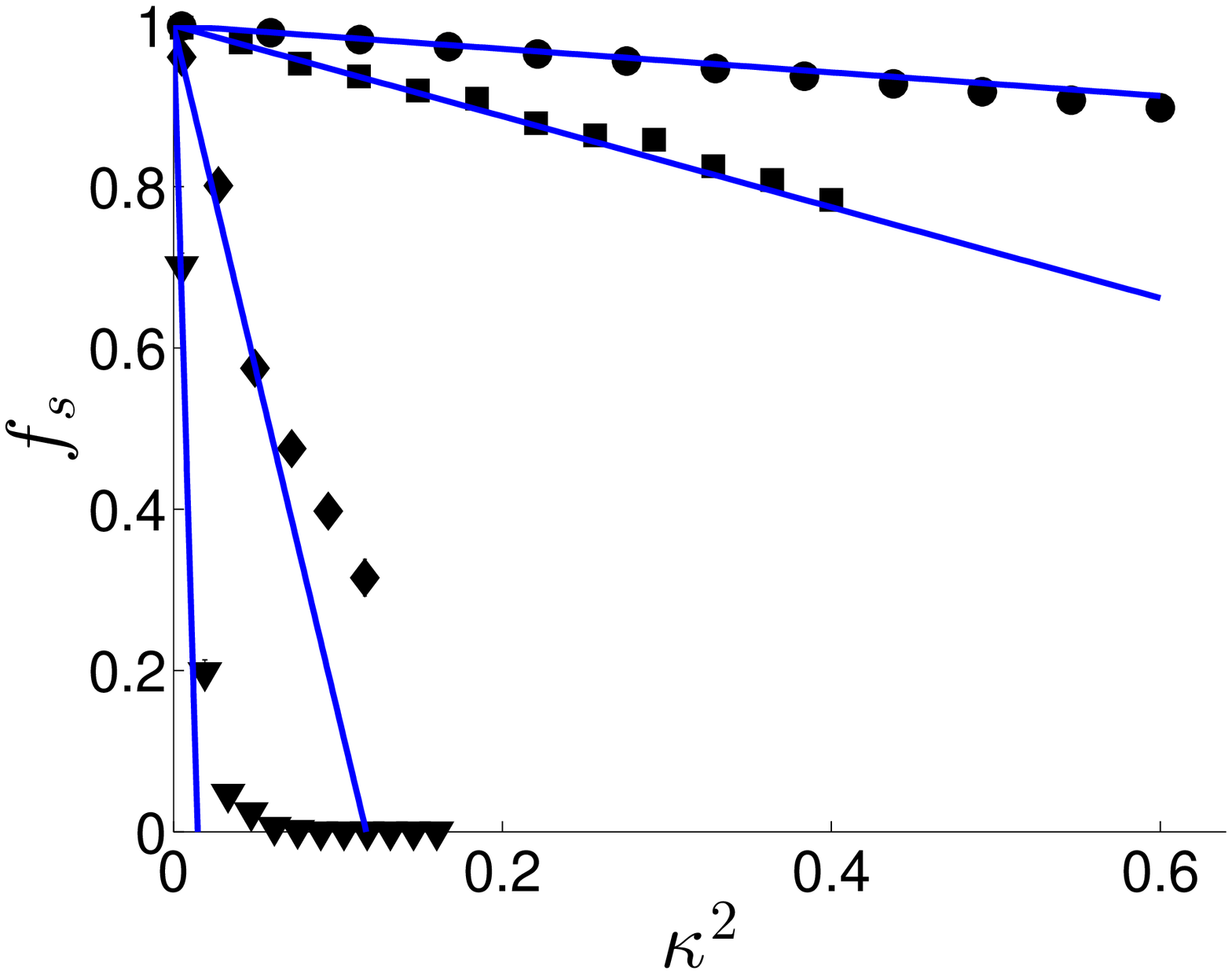}
	\caption{\label{fig:fsvskappa} (color online). Superfluid stiffness as a function of disorder strength, $\kappa^2$, for non-equilibrium parameters $\alpha=0.1,0.3,0.6,1$ from highest to lowest curves, respectively, and system sizes $L=64$ (left panel) and $96$ (right panel). Points show numerical results, and lines the perturbative expression, \er{eq:fs}. Numerical results are averages over 120 disorder realizations.}
\end{figure}
To go beyond  perturbation theory, we solve the eGPE numerically  on a discrete lattice of spacing $a_L=\xi$. At each site,  
the potential is independently drawn from a  Gaussian distribution of variance $\kappa^2$ 
by using a Mersenne Twister generator. Starting from a spatially constant density and phase we evolve the eGPE until a steady-state is reached. In the parameter range studied the steady-state is stable against perturbing the initial state. For each disorder realization, we then apply twisted boundary conditions, $\theta \in [-1,1]$ and increasing $\theta^2$ in steps of $0.25$, and evolve the eGPE to find the perturbed steady-state. The resulting frequency response fits to a quadratic function of $\theta$, allowing us to extract the stiffness from \er{eq:stiffness}, and we finally average over disorder realizations.

Fig.~\ref{fig:fsvskappa} shows how the stiffness obtained numerically compares with Eq. \eqref{eq:fs}, for different system sizes and non-equilibrium parameters. We see that when the condensate remains stiff, $f_s\lesssim 1$, the perturbative result agrees both qualitatively and quantitatively with simulations. However, in the regime where the stiffness is strongly suppressed,  the decay of $f_s(\kappa)$ deviates from the analytical prediction, even if $\kappa\ll 1$. The suppression of superfluidity in this strong fluctuation regime is thus not accurately described by perturbation theory. Nonetheless, the divergent perturbative result suggests a mechanism controlled by $L/\Ls$ which we will confirm in the following by further numerical  investigations.

\begin{figure}[tb]
	\centering
	\includegraphics[width=0.48\hsize]{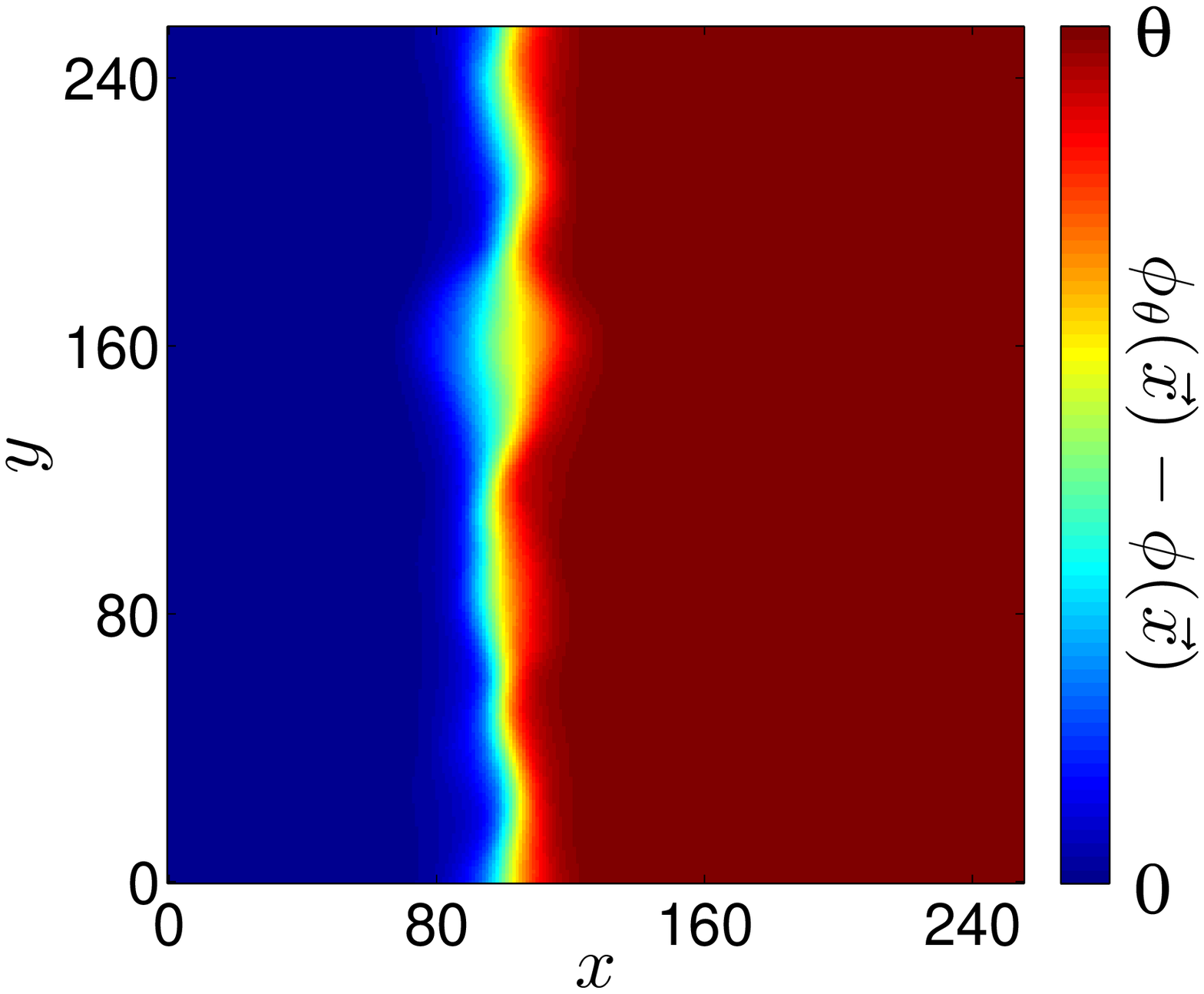}\nolinebreak \hfill
	\includegraphics[width=0.48\hsize]{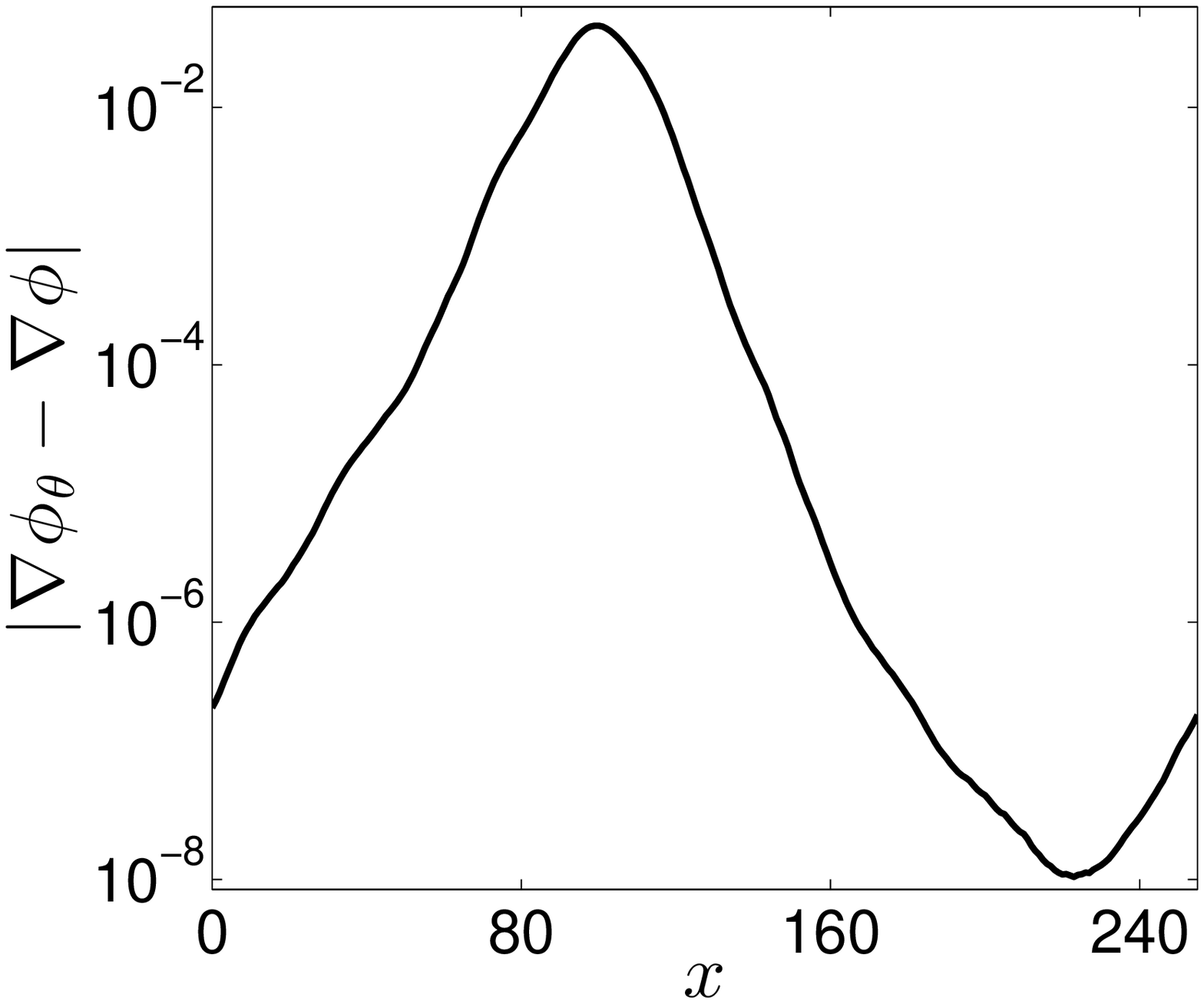}
	\caption{\label{fig:dwsnap} (color online). Phase (left) and current (right) response due to a phase twist $\theta$ along $x$, in a typical disorder realization. The plotted current response is $\nabla \phi_\theta(x) - \nabla \phi(x)$ averaged along $y$. Note the exponentially decaying tails of the current, c.f.~\er{eq:ansatzdw}, and the formation of a domain wall in the phase. Parameters used are $\alpha = 0.5$, $\kappa = 0.5$, $\theta = 1$.}
\end{figure}
\begin{figure}[tb]
	\centering
	\includegraphics[width=0.48\hsize]{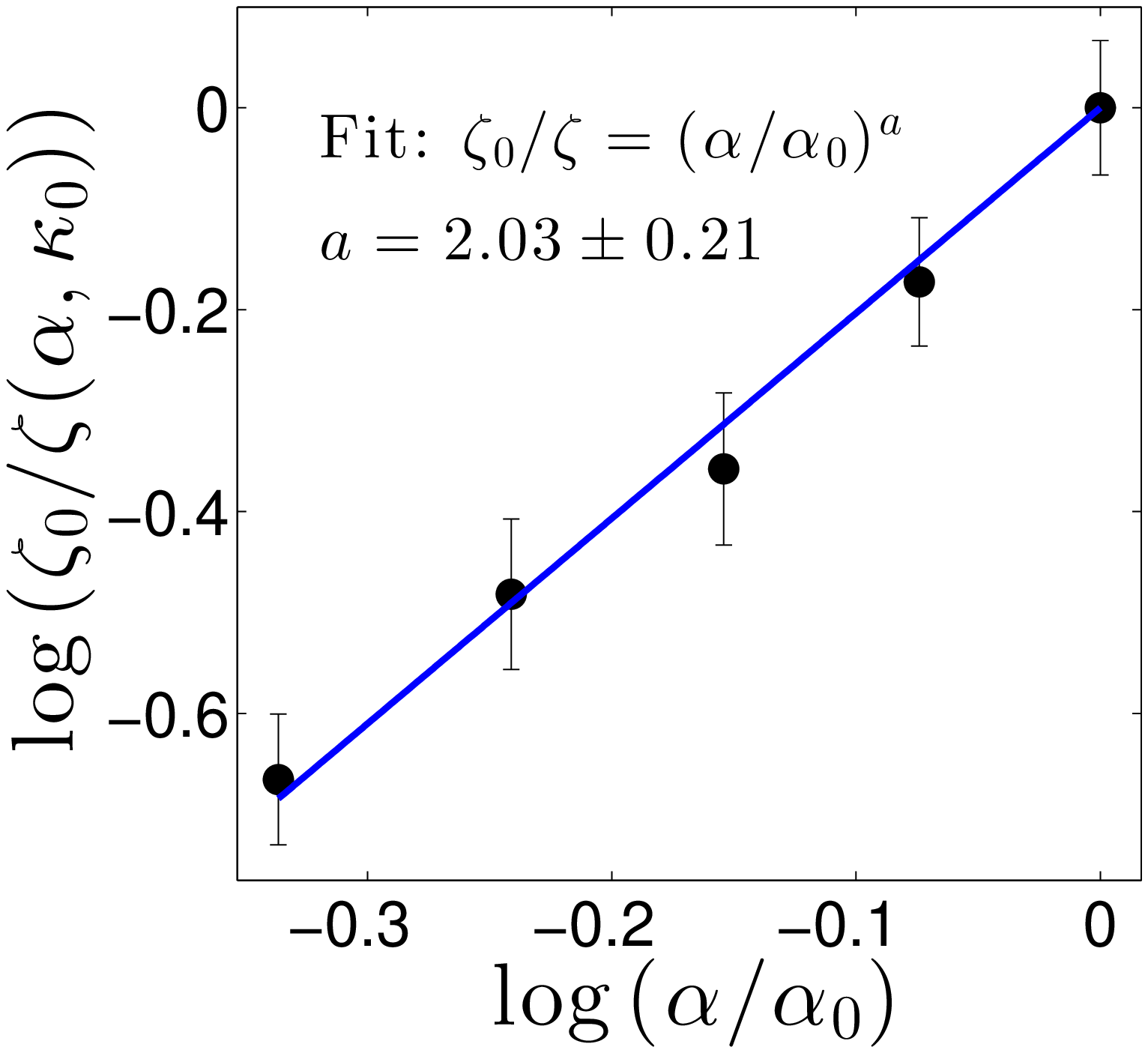}\nolinebreak \hfill
	\includegraphics[width=0.48\hsize]{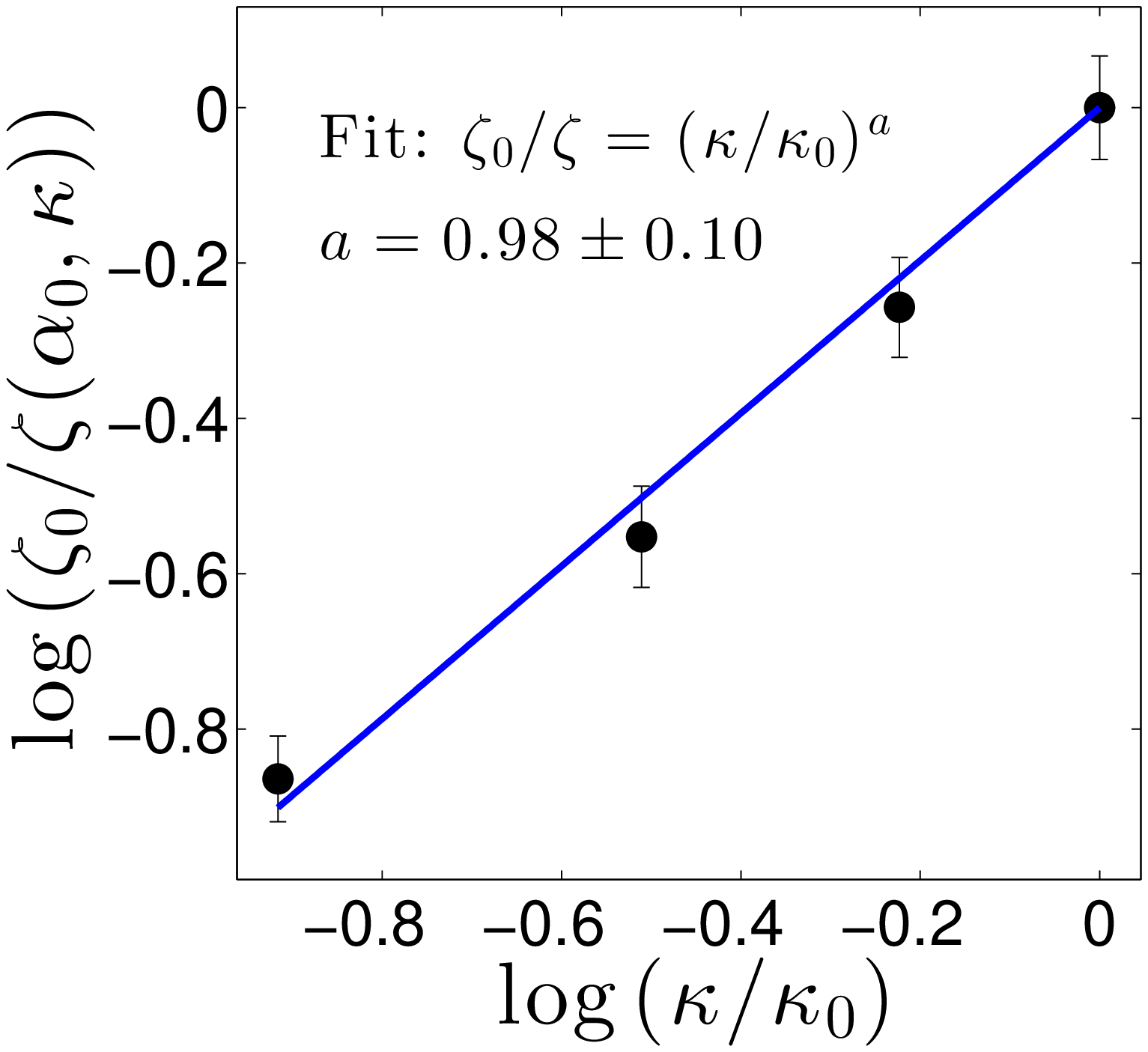}
	\caption{\label{fig:dwscaling} (color online). Scaling behavior of the decay length $\zeta$ with $\alpha$ and $\kappa$. Points are disorder averaged numerical results for the inverse of $\zeta(\alpha,\kappa)$ normalized to a reference value $\zeta_0 \equiv \zeta (\alpha_0,\kappa_0)$. Lines are linear fits on a double log scale. Left: dependence on $\alpha$ at fixed $\kappa = \kappa_0$ shows $\zeta^{-1}\sim\alpha^{2}$.  Right: dependence on $\kappa$ at fixed $\alpha = \alpha_0$ shows $\zeta^{-1}\sim\kappa$.  The parameters are $L = 256\: a_L$, $\alpha_0 = 0.7$, $\kappa_0 = 0.25$, $\theta = 1$, 72 disorder realizations.}
\end{figure}

To this end we have studied two limiting cases. First, if $L \ll \Ls$ the response to the twist is almost homogeneous, $\nabla \phi_ \theta -\nabla \phi \approx \twist$, and well described by perturbation theory, c.f. the discussion above. Second, if $L \gg \Ls$ the phase response occurs in two domains, with $\phi_\theta - \phi \approx 0$ and $\phi_ \theta - \phi \approx \theta$, separated by a randomly pinned domain wall of thickness $\sim \Ls$. This behavior is shown in Fig.~\ref{fig:dwsnap}. The associated density response (not shown) involves the left (right) edge of the domain wall forming a source (sink), as described by \er{eq:hydro2}. This allows a current response that is localized inside the domain wall. These results motivate the ansatz
\begin{align}
	\label{eq:ansatzdw}
	\nabla \phi_\theta - \nabla \phi = \frac{\theta}{2 \zeta (1-e^{-\frac{L}{2\zeta}})}   e^{-\frac{|x-x_0|}{\zeta}} \vec{e}_ \theta \ ,
\end{align}
where $x_0$ denotes the domain wall center, $\zeta$ the domain wall
size, and the amplitude is fixed by the twist angle. We extract
$\zeta$ from simulations by fitting to \er{eq:ansatzdw} in each
disorder realization before averaging. As discussed above, in
perturbation theory $\Ls$ is the relevant length scale, suggesting
that $\zeta\sim\Ls\sim1/\alpha^2\kappa$. This scaling is confirmed by
our simulations, as shown in Fig.~\ref{fig:dwscaling}.  In the
parameter range used we have observed single domain walls,
only. However, the formation of several walls might be
possible. Nevertheless, the identified mechanism relies on the
localization of the response, which remains present for several domain
walls.

\begin{figure}[tb]
	\centering
	\includegraphics[width=0.8\hsize]{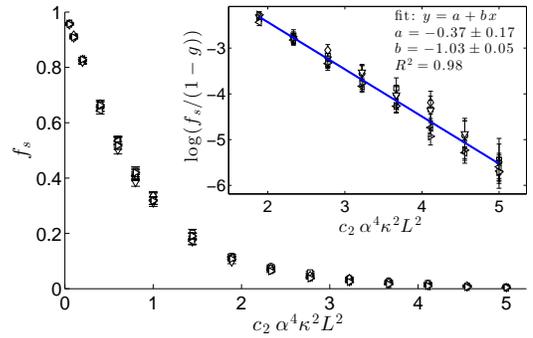}
	\caption{\label{fig:fsscaling} (color online). The numerically calculated stiffness as a function of $c_2 \alpha^4 \kappa^2 L^2 \sim L^2/\Ls^2$ shows a clear data collapse. Inset: comparison between the exponential tail and the scaling form,~\er{eq:fsscaling}, using the values of $c_2$ and $g$ obtained perturbatively (details see text). Points are shown for $L=64$ and $96$; $\alpha=0.9,1$ and $1.2$. For each data point we simulated up to 1320 disorder realizations.\vspace{-0.2in}}
\end{figure}

We also propose a scaling ansatz for the stiffness
\begin{align}
	\label{eq:fsscaling}
	f_s = e^{-c_2\,\alpha^4\kappa^2L^2} (1-g(\alpha,\kappa,\log L)),
\end{align}
to generalize the perturbative result into the regime of vanishing stiffness. This reproduces ~\er{eq:fs} when the exponential is expanded to first order, and takes into account that the dominant mechanism suppressing the stiffness is controlled by $L/\Ls$. The function $g$ includes logarithmic corrections and the equilibrium result, and at lowest order in perturbation theory is $g=(c_1+g_1(L)\alpha^2+g_2(L)\alpha^4)\kappa^2$, c.f. Eq. \eqref{eq:fs}. The simulation results, shown in Fig~\ref{fig:fsscaling}, confirm a clear data collapse with $\alpha^2\kappa L\sim L/\Ls$.  The exponential behavior in the regime $c_2\, \alpha^4\kappa^2L^2 \gtrsim 1$ is   in very good agreement with the scaling form incorporating the perturbative results for $c_2$, as shown in the inset. Note that to compare with simulations we calculate the perturbative form retaining the sums over discrete wavevectors; with this infrared regularization $c_2=7.734 \times 10^{-3}$.
%

%%% Experimental proposal %%%
%
\begin{figure}[b]
	\centering
	\includegraphics[width=0.65\hsize]{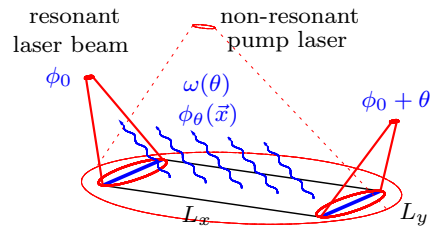}
	\caption{\label{fig:setup} (color online). Proposed measurement of condensate stiffness via the response of the condensate emission frequency $\omega$ or phase profile $\phi$ to a phase twist $\theta$ (see text).}
\end{figure}
Finally we propose an experiment, illustrated in Fig.~\ref{fig:setup}, to measure the superfluid stiffness of the non-equilibrium polariton condensate. We note that both the emission frequency and phase profile of the condensate can be measured~\cite{Kasprzak:2006,Roumpos:2012} while a phase twist could be imposed by driving with two coherent beams, resonant with the condensate, along either edge. In the limit of zero effective temperature, considered here, phase-locking~\cite{Wouters:2008,Eastham:2008,Pikovsky:2003} will pin the condensate phases at the boundaries to these beams and hence enforce a phase difference, $\theta$, across the condensate. Measuring the condensate emission frequency for various twists $\theta$, retuning the locking lasers appropriately, could allow the stiffness to be determined via~\er{eq:stiffness}. Alternatively, the phase map with the imposed phase twist $\phi_\theta(\vec{x})$, obtainable interferometrically, would show the characteristic formation of a domain wall, as in Fig. \ref{fig:dwsnap}, when compared with the untwisted case.

%%% Conclusion %%%
In conclusion, we have 
 found that the superfluid stiffness $f_s$ of a driven quantum condensate in a random potential vanishes in the thermodynamic limit for any non-zero disorder strength. In a finite system,  it decays exponentially with size, $f_s \sim e^{-(L/\Ls)^2}$, with the  length scale $\Ls\sim 1/\alpha^2\kappa$ controlling the decay. As $\Ls$  decreases when moving away from equilibrium or the clean limit, our work shows that the universal properties of driven condensates are completely different from those of equilibrium ones, if there is any static disorder. These predictions could be tested by measuring the phase profiles and emission frequency of  a polariton condensate in the presence of an imposed phase twist.

%%% Acknowledgement %%%
We thank A. Amo, S. Richter, R. Schmidt-Grund and M. Thunert for stimulating and helpful discussions concerning experiments. AJ is supported by the Leipzig School of Natural Sciences BuildMoNa, TH by the Dutch Science Foundation NWO/FOM and PRE by Science Foundation Ireland (09/SIRG/I1592).

%%% bibliography %%
% \bibliographystyle{apsrev4-1}
% \bibliography{bib_polariton}

%merlin.mbs apsrev4-1.bst 2010-07-25 4.21a (PWD, AO, DPC) hacked
%Control: key (0)
%Control: author (72) initials jnrlst
%Control: editor formatted (1) identically to author
%Control: production of article title (-1) disabled
%Control: page (0) single
%Control: year (1) truncated
%Control: production of eprint (0) enabled
%

\end{document}